\documentclass[aps,preprint,onecolumn,superscriptaddress,groupedaddress]{revtex4-2}
\usepackage{graphicx}  
\usepackage{dcolumn}   
\usepackage{bm}        
\usepackage{amssymb}   
\usepackage{upgreek}
\usepackage{mathtools}
\usepackage{hyperref}
\hypersetup{hidelinks}
\renewcommand{\hat}{\widehat}

\renewcommand{\i}{\mathrm{i}}
\renewcommand{\d}{\mathrm{d}}
\newcommand{\p}{\partial}
\newcommand{\h}{\hbar}
\newcommand{\e}{\mathrm{e}}
\renewcommand{\pi}{\uppi}

\renewcommand{\epsilon}{\varepsilon}

\begin{document}

\title{Holographic Approach to Neutron Stars}
\author{Tinglong~Feng}
\homepage{https://arendelle-ftl.github.io/}
\email{arendelle.ftl@gmail.com}
\affiliation{
	Xi'an Jiaotong University
}
\date{\today}
\begin{abstract}
In this article we explore the holographic approach to neutron stars in the realm of Quantum Field Theory (QFT). We delve into the structures of neutron stars, emphasizing the application of the AdS/CFT duality in modeling them. We discuss both "bottom-up" and "top-down" holographic models, comparing their predictions with astrophysical observations. Finally, we demonstrate the potential broader applications of the holography method in areas like superconductivity, highlighting the methodological significance of string theory and QFT in astrophysics.
\begin{description}
	\item[Key words]
	Holography, neutron stars, quantum chromodynamics
\end{description}
\end{abstract}

\maketitle

\section{Introduction}\label{I}
After a massive star experiences a supernova explosion and a subsequent gravitational collapse, a compact star emerges as a remnant. If the mass of the original hydrongen-burning star is not high enough to result in a direct collapse into a black hole, the remnant will be a neutron star, in which the Fermi and interaction repulsion should be able to withstand the gravitational collapse. A general method to deal with the corresponding strong nuclear force is Quantum Chromodynamics(QCD), which are often approached through weak-coupling expansion. Nevertheless, at neutron-star cores, QCD remains a strong coupled theory such that the perturbative methods fail, and we need nonperturbative method. \cite{pdg}

Holography is an impressive nonperturbative method based on famous AdS/CFT duality, which is a strong/weak duality in the sense that the gravitational theory is weakly curved when the effective coupling of the gauge is large, which means this method is suitable for describing some strongly interacting matter, such as neutron stars. 

In this article we will give a brief review of some novel holographic models for building a neutron star. In the next section, we will recall the basics equations for describing stars and neutron stars. In section \ref{III} we will give a brief review of QFT as necessary building blocks that enable studying holographic method.  In section \ref{IV} we will introduce AdS/CFT duality and discuss two holographic models, together with the predictions they give. Section \ref{V} is the summary of this article.
\section{From Stars to Neutron Stars}\label{II}
\subsection{Stars}
First, let's consider a star\cite{Weinberg_2019}. Suppose it is in equilibrium and is spherically symmetric, then a thin spherical shell feels a gravitational force
\begin{equation}
	F_1=-G\frac{4\pi r^2\rho(r) \d r M(r)}{r^2}=-4\pi G\rho(r)M(r)\d r
\end{equation}
where $M(r)$ is the total mass interior to $r$ satisfying
\begin{equation}\label{1.2}
\frac{\d M(r)}{\d r}=4\pi r^2\rho(r). 
\end{equation}
The buoyant force provided by pressure is
\begin{equation}
	F_2=4\pi r^2[p(r)-p(r+\d r)]=-4\pi r^2p'(r)\d r.
\end{equation}
For equilibrium condition we have $ F_1=-F_2$, hence
\begin{equation}\label{1.4}
	\frac{\d p(r)}{\d r}=-\frac{G M(r)\rho(r)}{r^2}.
\end{equation}
Equation \eqref{1.2} and \eqref{1.4} are hydrostatic equilibrium condition for stars.
\subsection{Neutron Stars}
Now let's turn to neutron stars. Obviously in a neutron star nuclear reactions have already ended so that its temperature tends to zero. Recall neutrons are Fermions, we could treat neutron stars as the ideal degenerate Fermi gas at $T=0$, a model with which we have been familiar when learning statistical mechanics in kindergarten. According to Fermi statistics the mass density could be denoted by 
\begin{equation}\label{1.5}
	\rho(r)=\frac{8\pi m_{\mathrm{n}}}{h^3}\int_0^{k_{\mathrm{F}}(r)}k^2\d k=\frac{8\pi m_{\mathrm{n}}k^3_{\mathrm{F}}}{3h^3}
\end{equation}
where $m_\mathrm{n}$ is the neutron mass and $k_\mathrm{F}$ is the Fermi momentum, the maximum momentum of the filled neutron levels. In a similar way we could obtain pressure 
\begin{equation}
	p(r)=\frac{8\pi c^2}{3h^3}\int_0^{k_{\mathrm{F}}(r)}\frac{k^4}{\sqrt{k^2c^2+m^2_{\mathrm{n}}c^4}}\d k.
\end{equation}
For a given central density $\rho(0)$, we could use \eqref{1.5} together with hydrostatic equilibrium \eqref{1.2},\eqref{1.4} to obtain $\rho(r)$, and then mass $M$ and radius $R$. Now we try to find  the maximum mass of neutron stars. Notice that there is a critical density $\rho_c$ where $k_{\mathrm{F}}=m_{\mathrm{n}}c $
	\begin{equation}
		\rho_c=\frac{8\pi m^4_{\mathrm{n}}c^3}{3h^3}=6.11\times10^{15}\mathrm{g/cm^3}
	\end{equation}
and the mean separation between neutrons
\begin{equation}\label{1.8}
	\left(\frac{\rho}{m_{\mathrm{n}}}\right)^{-1/3}=\left(\frac{\rho_c}{\rho}\right)^{1/3}\times0.52\times10^{-13}\mathrm{cm}.
\end{equation}
When we consider $\rho\geq\rho_c$, a problem arises that the neutron velocities are similar to $c$ such that we need to use general gravity to figure out the structure of neutron stars, calculations corresponding to which by Oppenheimer and Volkoff\cite{Op} suggested the maximum mass is $0.7M_\odot$ (where $M_\odot$ is solar mass). Nevertheless this calculations are also inaccurate, for if $\rho\geq\rho_c$, \eqref{1.8} shows that the seperation of neutrons is less than the range of nuclear force such that we can no longer treat neutrons as ideal Fermi gas. Actually we have found neutron stars with $2.1M_\odot$\cite{Cromartie2020} or even $2.5M_\odot$\cite{Abbott_2020} masses, according to recent observations. Hence we must use QCD to treat this situation. Nevertheless, as we have said in section \ref{I}, perturbative QCD can not be used in this situations as well, so we must turn to nonperturbative method, for example, holography.
\section{A Brief Review of QFT}\label{III}
Before entering holographic method, we shall first give a short review\cite{Weinberg_1995} of QFT to provide a necessary foundation.
\subsection{Canonical Formalism}
The canonical quantization of a field obeys following procedure
\begin{enumerate}
	\item writing down the Lagrangian density $\mathcal{L}(\phi,\p_\mu\phi)$ of the field, using the Euler-Lagrange equation" to obtain field equation
	\item figuring out the canonical momentum density $\Pi^\mu=\p\mathcal{L}/\p(\p_\mu\phi)$ and Hamiltonian density $\mathcal{H}=\Pi^0\p_0\phi-\mathcal{L}$
	\item Transferring the field and momentum density to operators satisfying canonical commutation relations
	\item Using creation and annihilation operators to express field operator
\end{enumerate}
For example, let's quantize a free real scalar field (we will use natural units below). The Lagrangian density is 
\[ \mathcal{L}=\frac{1}{2}\p_\mu\phi(x)\p^\mu\phi(x)-\frac{1}{2}m^2\phi(x)^2 \]
canonical momentum density
\[ \Pi^0(x)=\p^0\phi(x) \]
Hamiltonian 
\[ \mathcal{H}=\frac{1}{2}[\p_0\phi(x)]^2+\frac{1}{2}[\bm{\nabla}\phi(x)]^2+\frac{1}{2}m^2\phi^2(x) \]
Transfer field and momentum density to operators
\[ \phi(x)\to\hat{\phi}(x),\quad \Pi^0(x)\to\hat{\Pi}^0(x)\]
satisfying
\[ [\hat{\phi}(t,\bm{x}),\hat{\Pi}^0(t,\bm{y})]=\i\bm{\delta}(\bm{x}-\bm{y}),\]
\[ [\hat{\phi}(\bm{x}),\hat{\phi}(\bm{y})]=[\hat{\Pi}^0(\bm{x}),\hat{\Pi}^0(\bm{y})]=0 \]
Use creation and annihilation operators to express field operator
\[ \hat{\phi}(\bm{x})=\int\frac{\d^3p}{\sqrt{(2\pi)^3 2E_{\bm{p}}}}\left(a_{\bm{p}}\e^{-\i(E_{\bm{p}} t-\bm{p}\cdot\bm{x})} +a^\dag_{\bm{p}}\e^{\i(E_{\bm{p}} t-\bm{p}\cdot\bm{x})} \right) \]
where
\[ E_{\bm{p}}=\sqrt{\bm{p}^2+m^2} \]
\subsection{Path-Integral Methods}
There's another method besides canonical formalism, i.e. the path-integral method, which will be useful in AdS/CFT. In quantum mechanics we have sum over path\cite{Feynman}
\begin{equation}
	K(a,b)=\int_{a}^{b}\mathcal{D}x(t)\e^{(\i/\h)S[b,a]}
\end{equation}
Similarly, in QFT we have generating functional 
\begin{equation}\label{10}
	Z[J]=\int\mathcal{D}[\phi(x)]\exp\left\{\i\left[S+\int J(x)\phi(x)\right] \right\}
\end{equation}
where $\phi(x)$ is the field, $S$ is the action, $J(x)$ is a source. From the generating function we can obtain information the quantum field, such as Green's function.
\section{Holographic Neutron Stars}\label{IV}
\subsection{AdS/CFT duality}

Holography, or AdS/CFT duality, provides the generating fucntional in $D$-dimensional Minkowski space by using classical action $S[\phi^{\mathrm{AdS}}]$ of a field on (at least) $D+1$-dimensional Anti-deSitter Space\cite{AdS}. 
\begin{figure}[h]
	\includegraphics[width=0.5\textwidth]{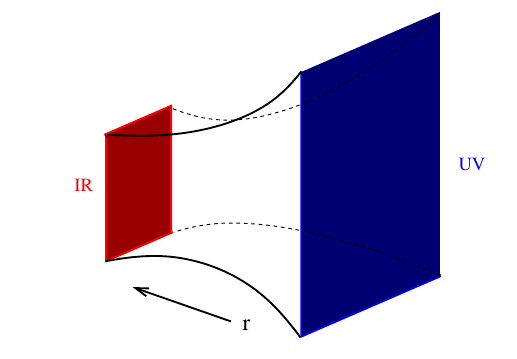}
	\caption{\label{fig1} AdS space, taken from\cite{DTong}}
\end{figure}
Fig \ref{fig1} shows AdS/CFT, where the right blue plane representing the Minkowski space the QFT lives in, often called the boundary. Leaving away from the boundary is another dimension, labelled with $r$. The larger space usually called the bulk, inside which a theory of gravity lives. Holography claims that gravity in the bulk and QFT on the boundry are equivalent\cite{DTong}.  The fundamental formula\cite{Wit} of hologrophy is 
\begin{equation}
	Z_{\mathrm{QFT}}[\phi_0]=Z_{\mathrm{QG}}[\phi]|_{\phi\to\phi_0}
\end{equation}
where QG denotes quantum gravity, and $Z_{\mathrm{QFT}}[\phi_0] $ is indeed from \eqref{10}. The most impressive feature is, when the QFT has a large number of degrees of freedom and is strongly coupled, the dual description reduces to classical gravity\cite{Mal}, hence
\begin{equation}\label{12}
	Z_{\mathrm{QFT}}[\phi_0]\simeq\e^{\i S_{\mathrm{bulk}}}|_{\phi\to\phi_0}.
\end{equation}
\eqref{12} tells us that if the classical action $S_{\mathrm{bulk}}$ is given, we can calculate its dual description in the boundry, i.e. , a strongly coupled QFT, such as strongly coupled QCD. 
\subsection{Building a Realistic Neutron Stars from Holography}
Generally speaking, there are two kinds of holographic models for QCD: the "bottom-up" models and the "top-down" models. The former are more phenomenological while the latter are rigorously based on string theory. Both of them can be used to build a realistic neutron stars which provide meaningful predictions. We will discuss two novel models, one from each kind, and review their predictions.

The first one\cite{M1-1,M1-2} employs a "top-down" approach, the Saki-sugimoto model, to build a neutron star, which has two parameters, the 't Hooft coupling $\lambda$ and the Kaluza-Klein mass $M_{\mathrm{KK}}$. Adding leptons, and constructing a holographic crust in order to account for realistic neutron stars, the model figures out the scales of mass and radius
\begin{equation}
	M_0\simeq1.445\lambda_0^{-3/2}\left(\frac{M_{\mathrm{KK}}}{\mathrm{GeV}}\right)^{-2}M_\odot
\end{equation}
\begin{equation}
	r_0\simeq2.135\lambda_0^{-3/2}\left(\frac{M_{\mathrm{KK}}}{\mathrm{GeV}}\right)^{-2}\mathrm{km}
\end{equation}
\begin{figure}[h]
	\includegraphics[width=0.5\textwidth]{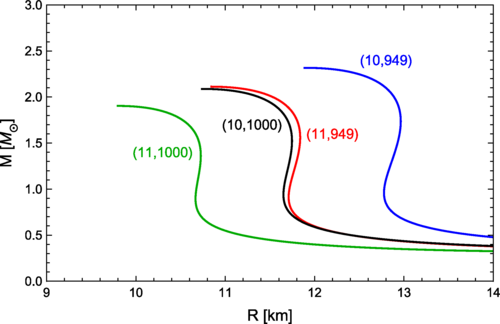}
	\caption{\label{fig2} Mass-radius curves for different parameter sets, taken from\cite{M1-1}}
\end{figure}
Figure \ref{fig2} shows the Mass-radius curves obtained in this model. Besides that, this model also obtains tidal deformability using numerical method.

The second one is a "bottom-up" model\cite{M2-1,M2-2}, inspired by holographic electric superconductors. The action is 
\begin{equation}
	S_{\mathrm{Bulk}}+\int\d^{d+1}x\sqrt{-g}\left\{\mathcal{R}+\frac{d(d-1)}{L^2}-\frac{1}{4}F^2\right\}.
\end{equation}
The equation of state can be derived from this model
\begin{equation}
	p=\epsilon-\sqrt{a\epsilon}\mu_c
\end{equation}
where $p$ is pressure, $\epsilon$ is energy density and $a,\mu_c$ are parameters. Together with TOV equations the $M(R)$ can be plotted as
\begin{figure}[h]
	\includegraphics[width=0.5\textwidth]{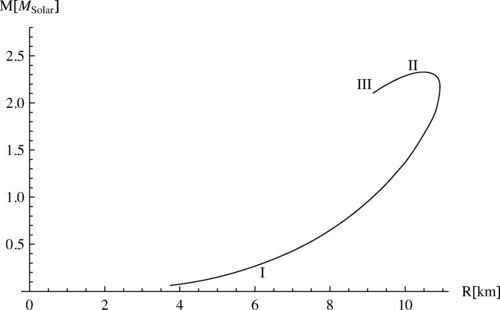}
	\caption{\label{fig3} Plot of $M(R)$, taken from\cite{M2-1}}
\end{figure}
\subsection{Prediction}
Both models mentioned above give predictions. The first one forecasts the maximal mass, radius and tidal deformability of corresponding neutron stars, as in Table \ref{t-1}. Among this, the upper limit of maximal mass, $2.46M\odot$, and the lower limit of tidal deformability of a $1.4M\odot$ neutron star, 277, are novel and independent prediction which could be verified according to future obsevational results.
\begin{table}[h]
	\caption{\label{t-1}Constraints obtained by combining the holographic results with astrophysical data\cite{1,2,3,4} for maximal mass as well as radius and tidal deformability $\Lambda$ in the "top-down" model }
	\begin{ruledtabular}
		\begin{tabular}{lcr}
			 & lower bound&upper bound\\
			\hline
			$M_{\max}[M\odot]$ & 2.1 & 2.46\\
			$R_{1.4}[\mathrm{km}]$ & 11.9 & 14.3\\
			$R_{2.1}[\mathrm{km}]$ & 11.4 & 13.7\\
			$\Lambda_{1.4}$ & 277 & 580\\
			$\Lambda_{2.1}$ & 9.13 & 29.3\\
		\end{tabular}
	\end{ruledtabular}
\end{table}

The second model confirms the possibility of the existence of a neutron star with $(M,R) \sim (2M\odot, 10\mathrm{km})$. Furthermore, for the model is inspired by holographic electric superconductors, it is suitable to investigate whether a neutron star with a color superconducting (CSC) core. According to \cite{M2-2}, this case is impossible for the instability of corresponding stars.
\section{conclusion}\label{V}
In this article we give a brief review of holographic method for building neutron stars, display some novel models and corresponding predictions. Holographic neutron star is just a simple application of the impressive AdS/CFT duality, which is also very useful in other areas, such as superconductivity\cite{C1,C2,C3,C4}. The fantastic power of AdS/CFT is due to its Strong/Weak duality, which could transfer strongly coupled field problems to almost classical gravity problems, and the latter is much easier to solve. This method displays methodological value of string theory, or more generally, QFT. Even if we have not yet formulated a widely accepted, consistent theory that combines general relativity and quantum mechanics, the current progress in QFT still provides us with valuable methods in various application areas.
\bibliography{ref}
\end{document}